\documentclass{elsart}

\usepackage{setspace}

\date{\today}
\usepackage{graphicx}
\usepackage{amssymb}
\usepackage{subfigure}
\usepackage{lineno}

\newcommand{\gm}{$(g-2)$}  

\begin{document}


\begin{frontmatter}
\title{A Tungsten / Scintillating Fiber Electromagnetic Calorimeter Prototype for a High-Rate Muon
\gm\ Experiment}

\author{R. McNabb},
\author{J. Blackburn},
\author{J.D. Crnkovic},
\author{D.W.~Hertzog},
\author{B. Kiburg},
\author{J. Kunkle},
\author{E. Thorsland},
\author{D.M. Webber}

\address{Department of Physics, University of Illinois at
         Urbana-Champaign, Urbana, IL 61801, USA}
\author{K.R. Lynch}
\address{Department of Physics, Boston University, Boston, MA 02215,
         USA}

\vspace{1cm}

Contact person:  \address{D.W. Hertzog, Department of Physics, 469
Loomis Laboratory of Physics, University of Illinois at
Urbana-Champaign, 1110 West Green Street, Urbana, IL 61801-3080.
Tel.: 1-217-333-3988; Fax:  1-217-333-1215; Email:
hertzog@uiuc.edu}.

\begin{abstract}
  A compact and fast electromagnetic calorimeter prototype was designed,
  built, and tested in preparation for a next-generation, high-rate muon \gm\ experiment.
  It uses a simple assembly procedure: alternating layers of 0.5-mm-thick tungsten
  plates and 0.5-mm-diameter plastic scintillating fiber ribbons. This geometry leads to a
  detector having a calculated radiation
  length of 0.69~cm, a Moli\`{e}re radius of 1.73~cm, and a measured
  intrinsic sampling resolution term of $(11.8\pm1.1)\%/\sqrt{E({\rm GeV})}$,
  in the range 1.5 to 3.5~GeV. The construction procedure, test beam
  results, and GEANT-4 comparative simulations are described.

\end{abstract}


\end{frontmatter} 

\section{Introduction} 

In the early 1980's, the production of thin clad scintillating
fibers led to development of fiber-based electromagnetic
calorimeters. Early examples include the Omega inner
calorimeter~\cite{Omega} and the Jetset forward
calorimeter~\cite{Hertzog:1990}.  Both featured 1-mm-diameter fibers
embedded in thin, grooved lead foils (Pb/SciFi).  The fibers were
orientated nearly head-on with respect to the incoming beam. Because
fiber attenuation lengths exceed 1~m, applications involving long
Pb/SciFi ``bars'' are also practical. The Jetset gamma barrel veto
counter consisted of 24 trapezoidal 80-cm-long bars formed into a
continuous cylindrical structure. The side-on
calorimeters~\cite{Sedykh:2000} for the E821 muon \gm\ experiment
were also made of Pb/SciFi with a radial readout of a large block
into four summed photomultiplier tubes (PMTs). The KLOE
collaboration constructed the largest installation of fiber-based
calorimetry~\cite{KLOE}. The entire detector, in collider-geometry,
was based on very long Pb/SciFi bars. With special tooling and
careful assembly, they achieved excellent energy and timing
resolution.

This report details studies
of a tungsten-based scintillating fiber (W/SciFi) calorimeter
prototype having a 0.5-mm layer thicknesses. While the motivation is
toward a new high-rate muon \gm\ experiment, the generic properties
of this type of dense, fast-response detector are of interest for
other applications. The experimental conditions that motivate a
calorimeter of this type will first be discussed, followed by our
reports on beam and simulation tests of its performance.

In the Brookhaven muon \gm\ experiment~\cite{Bennett:2006},
3.1~GeV/$c$ polarized muons are injected and kicked onto orbits of a
highly uniform magnetic storage ring~\cite{Danby:2001}. As the muons
circulate the ring, their spins rotate faster than the cyclotron
frequency with a frequency difference proportional to the anomalous
magnetic moment. Parity violation leads to a greater number of
high-energy decay positrons being emitted in the direction of the
muon spin.  The positrons curl to the inside of the storage ring
where they are intercepted by one of the 24 electromagnetic
calorimeters, positioned adjacent to the storage ring aperture. The
rate of positrons exceeding a given energy threshold---typically
~2~GeV---is proportional to an exponential modulated with the \gm\
spin precession frequency.  Each storage ring ``fill'' is observed
for $\sim700~\mu$s; the instantaneous rate varies from several MHz
at the beginning of the fill to tens of Hz at the end of the fill.
Because uncontrolled pileup of two nearly simultaneous events is a
severe systematic error for the experiment, the calorimeter response
must be capable of resolving two events separated by as little as
5~ns. Gain over the fill period must also remain very stable so that
the threshold instability does not affect the fitted frequency.

The electromagnetic calorimeters for the E821 experiment used a
Pb/SciFi design~\cite{Hertzog:1990,Sedykh:2000}. Each calorimeter
consisted of a monolithic block of 1-mm diameter fibers arranged in
a near close-packed geometry within grooved lead alloy foils. The
fractional composition of the detector was Pb:Sb:Fiber:Glue = $0.466
: 0.049 : 0.369 :  0.096$ (by volume), leading to a radiation length
$X_0 = 1.14$~cm.  The fibers were oriented radially so that the
positrons would impact on the detector at large angles with respect
to the fiber axis.  Four lightguides directed the light to
independent PMTs and the summed analog signal was processed by
waveform digitizers. The 14-cm high by 22.5-cm radial by 15-cm deep
calorimeter dimensions were largely dictated by the available space
and the need to have a sufficient radial extension to intercept the
positrons. The energy resolution requirement for \gm\ is relatively
modest, $\sim10\%$ or better at 2~GeV.

A next-generation muon \gm\ experiment~\cite{E969}---with an aim of
greater than four times improved precision---will require
$\sim20$-fold increase in total stored muons. The muon injection
rate will increase in most of the considered scenarios. This will
lead to a higher instantaneous rate on the calorimeters, which will
make the control of gain stability more difficult and will increase
the magnitude of the pileup correction. The systematic errors
associated with gain instability and pileup must each be reduced by
a factor of $\sim 3$. Accordingly, we have begun designing a new
calorimeter that retains the fast response time of plastic
scintillating fiber, but is made from an array of dense submodules
where each is oriented roughly tangential to the muon orbit. This
configuration will provide transverse segmentation and allow for
multiple simultaneous shower identification.  A 50:50 ratio of
tungsten to scintillator (and epoxy) reduces shower transverse and
longitudinal dimensions. The calculated~\cite{fabjan} radiation
length, $X_0 = 0.69$~cm, is $60\%$ of the length for the Pb/SciFi
modules used in E821. Consequently, the modules can be made compact
enough to free space for downstream readout in the highly
constricted environment of the storage ring. The high density leads
to a smaller radial shower size, which improves the isolation of
simultaneous events. We find that using 0.5-mm layers gives an
acceptable resolution close to $10\%$ at 2~GeV for our prototype; a
non-trivial error contribution to this performance parameter arises
from beam momentum spread, photo-electron yield and transverse
leakage fluctuations. Therefore, the intrinsic detector response
from sampling fluctuations alone is better.

\begin{figure}
\begin{centering}
\subfigure[Prototype module and guide]
{\includegraphics[width=0.45\textwidth]{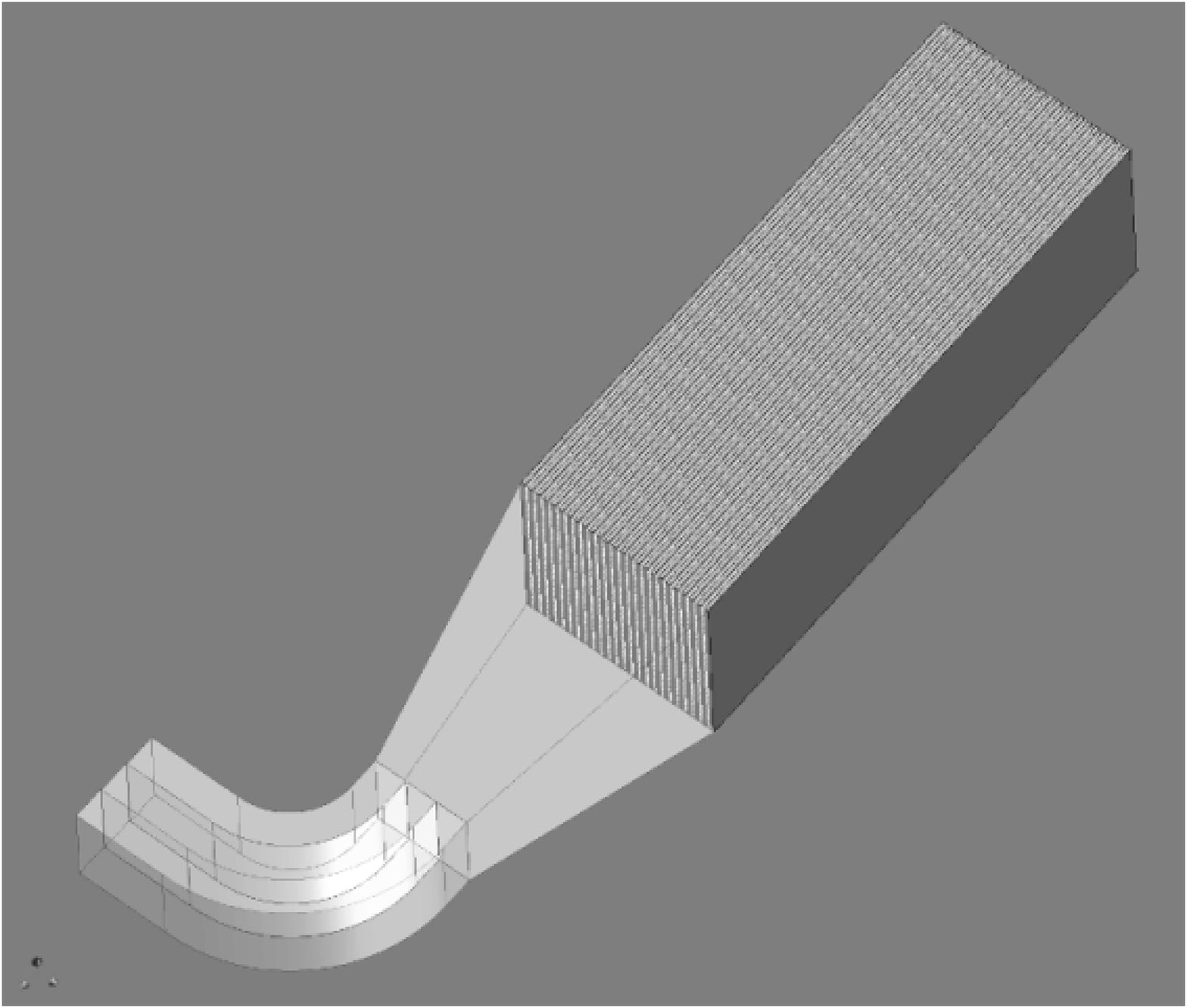}}
\subfigure[Proposed array of 20 modules]
{\includegraphics[width=0.45\textwidth]{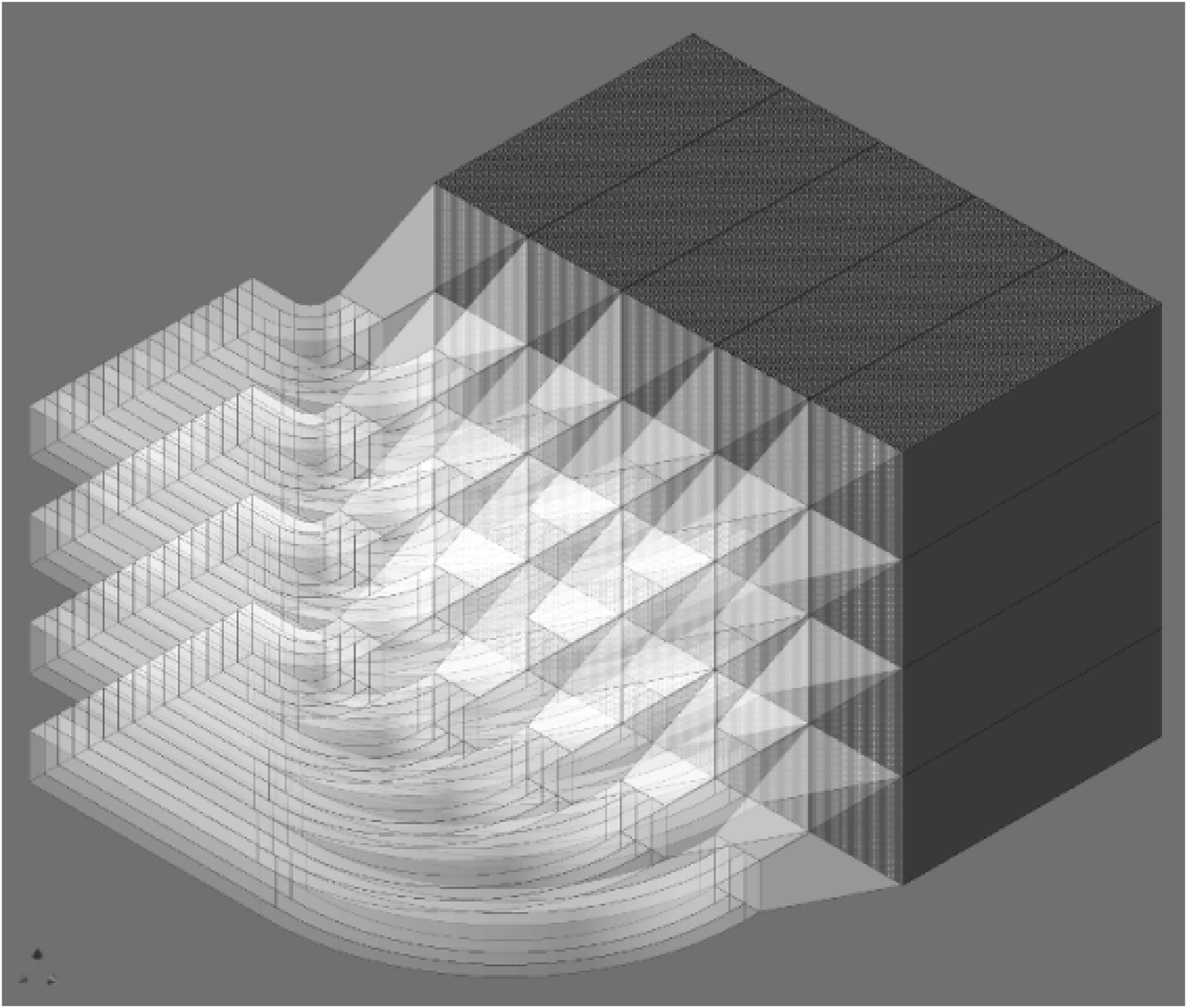}}

\end{centering}
\caption{a) Prototype $4 \times 6 \times 17$~cm$^3$ module and b)
proposed array of twenty $4 \times 4 \times 11$~cm$^3$ modules. The
lightguides must curl toward the \gm\ storage ring center and
connect to PMTs outside of the storage ring magnetic field.}
\label{fig:design}
\end{figure}

The $4\times6\times17$~cm$^3$ prototype is shown in
Fig.~\ref{fig:design}a.  The full conceptual design array of
$4\times4\times11$~cm$^3$ modules is shown in
Fig.~\ref{fig:design}b; 24 such arrays are required for the proposed
muon \gm\ experiment.   The depicted lightguides in
Fig.~\ref{fig:design}b imply a subdivision of the W/SciFi block into
20 independent readouts by PMTs (not shown), which must be placed
sufficiently far away so as to minimize effects on (and from) the
1.45~T precision storage ring magnetic field. New, large dimension
multi-pixel avalanche photodiodes, which are operated in Geiger mode
(a silicon photomultiplier or SiPM), are another option for readout.
These devices are compatible with placement in the field and, owing
to their smaller photo-sensitive surfaces, would likely lead to a
much greater segmentation on the downstream side of the calorimeter.

\section{Assembly}

One clear advantage of the present W/SciFi design over our previous
scintillating fiber calorimeters is the simplicity of the assembly
procedure. Tungsten is a very difficult material to machine, mold,
or extrude, so precisely rolled thin plates dimensioned to the size
of individual absorbers are used. While a module dimension is
envisioned to be $4\times4\times11$~cm$^3$, each of the 24
calorimeter stations can be built from a continuous W/SciFi block; a
single 16-cm high by 20-cm wide block can be attached to an array of
lightguide reducers on the downstream face. This will minimize the
segmentation boundaries, similar to the E821~\cite{Sedykh:2000}
four-lightguide design that attaches to the Pb/SciFi detector.

Our prototype module is made large enough to study the basic
properties of the envisioned calorimeter design, but with a
calculated Moli\`{e}re radius of $\sim1.73$~cm, it does not provide
full shower containment, even for centered events.  The calorimeter
components include a set of 0.5-mm thick $99.95\%$ pure tungsten
plates~\cite{tungstenplates} ($\rho = 19.0$~g/cm$^3$) measuring
$6.0\times16.5$~cm$^2$.  The delivered plates were all measured for
thickness at eight different points and the sample variance around
the mean of 0.51~mm was found to be $\pm0.04$~mm. The plate edges
were slightly out of square owing to the rolling procedure.  The
preliminary step of grinding the plate edges, using a magnetic
surface grinder with a silicon carbide wheel, was taken to produce
straight sides and 90-degree corners.

The 0.5-mm diameter BCF-20, ``green-emitting'' scintillating fibers
were obtained from Saint-Gobain Crystals~\cite{saintgobain}.  These
fibers were conveniently available owing to a large production for
an independent project.  They arrived as 12-cm wide by 17.5-cm long
``ribbons.'' Each ribbon came as a self-contained structure with the
individual 0.5-mm fibers held adjacent by a coating of a
polyurethane-acrylic blend cement. We split each ribbon into two
6-cm wide strips to match the tungsten plate widths. The fibers are
coated with a $10-15~\mu$m thick white extra mural absorber for
better light transmission. For the final production modules, BCF-10
(or equivalent) ``blue'' fibers will be used to better match the
quantum efficiency of the readout device and for faster time
response, but the green fibers were adequate for our basic tests.

\begin{figure}
\begin{centering}
\subfigure[Stacking plates and ribbons]
{\includegraphics[width=0.45\textwidth]{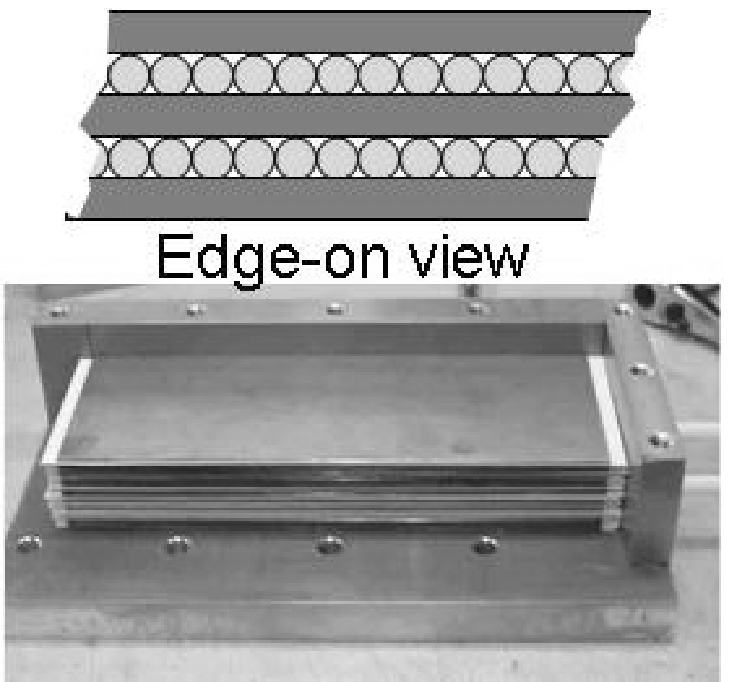}}
\subfigure[Finished stacking, before glue]
{\includegraphics[width=0.45\textwidth]{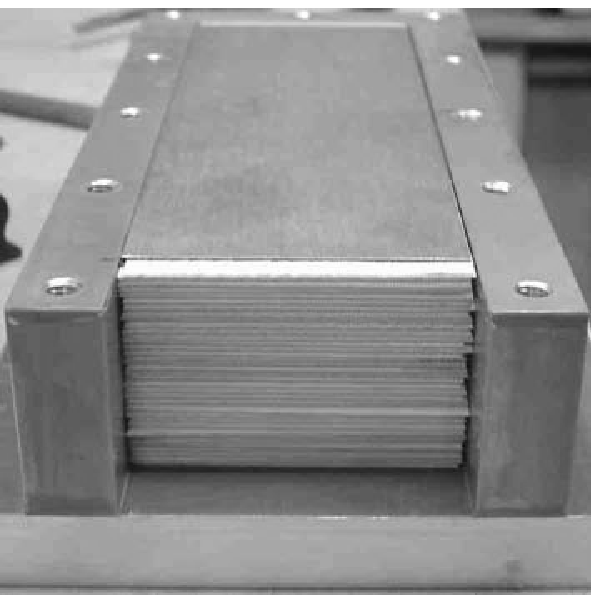}}
\subfigure[Machining ends]
{\includegraphics[width=0.45\textwidth]{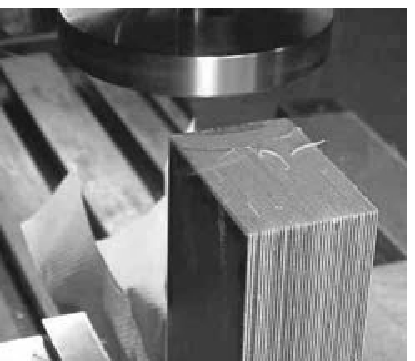}}
\subfigure[Illuminated face]
{\includegraphics[width=0.45\textwidth]{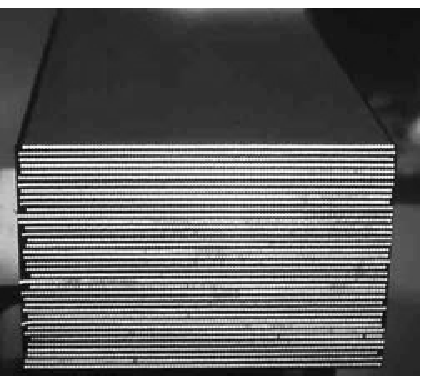}}

\end{centering}
\caption{The diagram and four photos represent components and steps
in the assembly process. a) Top: Edge-on view of the 0.5-mm fiber
ribbons between 0.5-mm-thick tungsten sheets. Bottom: Side view of
stacking actual plates and fibers; b) The completed stack without
the top jig plate; c) Machining off one of the end faces; d) Front
face, illuminated from rear. } \label{fig:assembly}
\end{figure}

The concept of plate / ribbon stacking is shown in the upper half of
Fig.~\ref{fig:assembly}a.  Assembly consisted of alternately
stacking tungsten plates and fiber ribbons in a jig, the lower half
of Fig.~\ref{fig:assembly}a and also Fig.~\ref{fig:assembly}b.  The
fiber ends were aligned with the right-side jig face, while the
tungsten plates were positioned roughly 0.5~cm from this jig edge.
After the stack was completed, Fig.~\ref{fig:assembly}b, a top and
side plate were attached, establishing the final fixed dimension.
The assembly was then turned upward and 1266 Stycast epoxy was
poured into the assembly in a manner to ensure a careful and uniform
distribution among all layers.  A vacuum was pulled from the bottom
through one of the tubes seen in Fig.~\ref{fig:assembly}a.  Once the
glue was drawn through the stack, being visible in the tube, the
tubing was removed and the two drainage holes on the right-side jig
face were plugged. Since the fibers extended beyond the tungsten
plates on both sides, an epoxy-fiber solid appendage was formed,
which could then be machined easily to a smooth and polished
surface, see Fig.~\ref{fig:assembly}c.  The final array, illuminated
by a standard lamp, is shown in Fig.~\ref{fig:assembly}d.

\section{Experimental Tests and Simulations} 

The W/SciFi detector was first tested at electron energies of 0.140,
0.245, and 0.300 GeV at the piM1 beamline at the Paul Scherrer
Institut (PSI), and later at the more relevant energies from 1.5 to
3.5~GeV at the Meson Test Beam at Fermilab (FNAL).  The focus of the
test beam measurements was on calorimeter linearity and energy
resolution.  While neither beamline was optimized to provide a small
momentum resolution or spot size, sufficient performance information
was obtained to compare measurement to GEANT-4 based Monte Carlo
simulations.

\begin{figure}
\begin{centering}
\subfigure[Data]
{\includegraphics[width=0.45\textwidth]{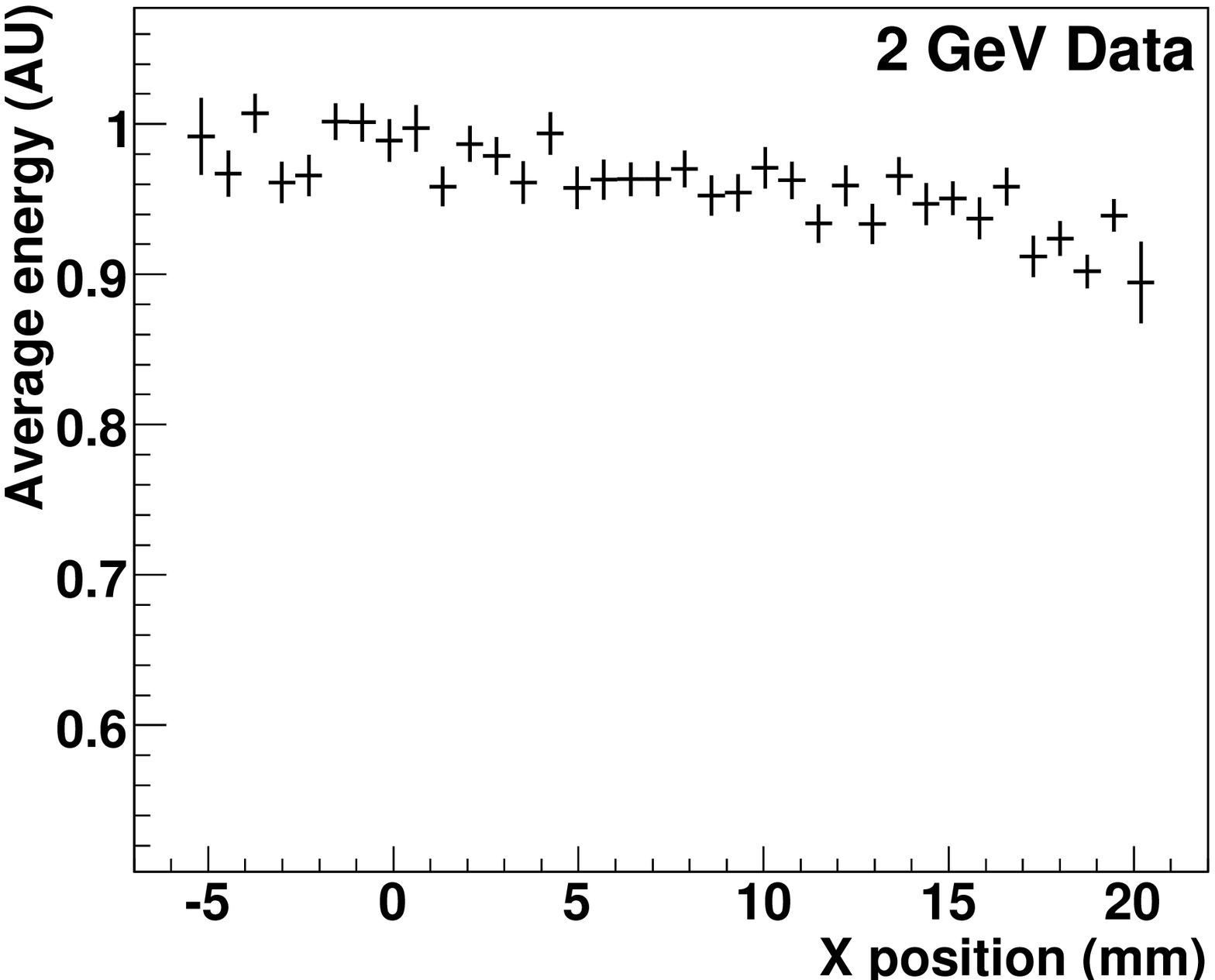}}
\subfigure[Monte Carlo]
{\includegraphics[width=0.45\textwidth]{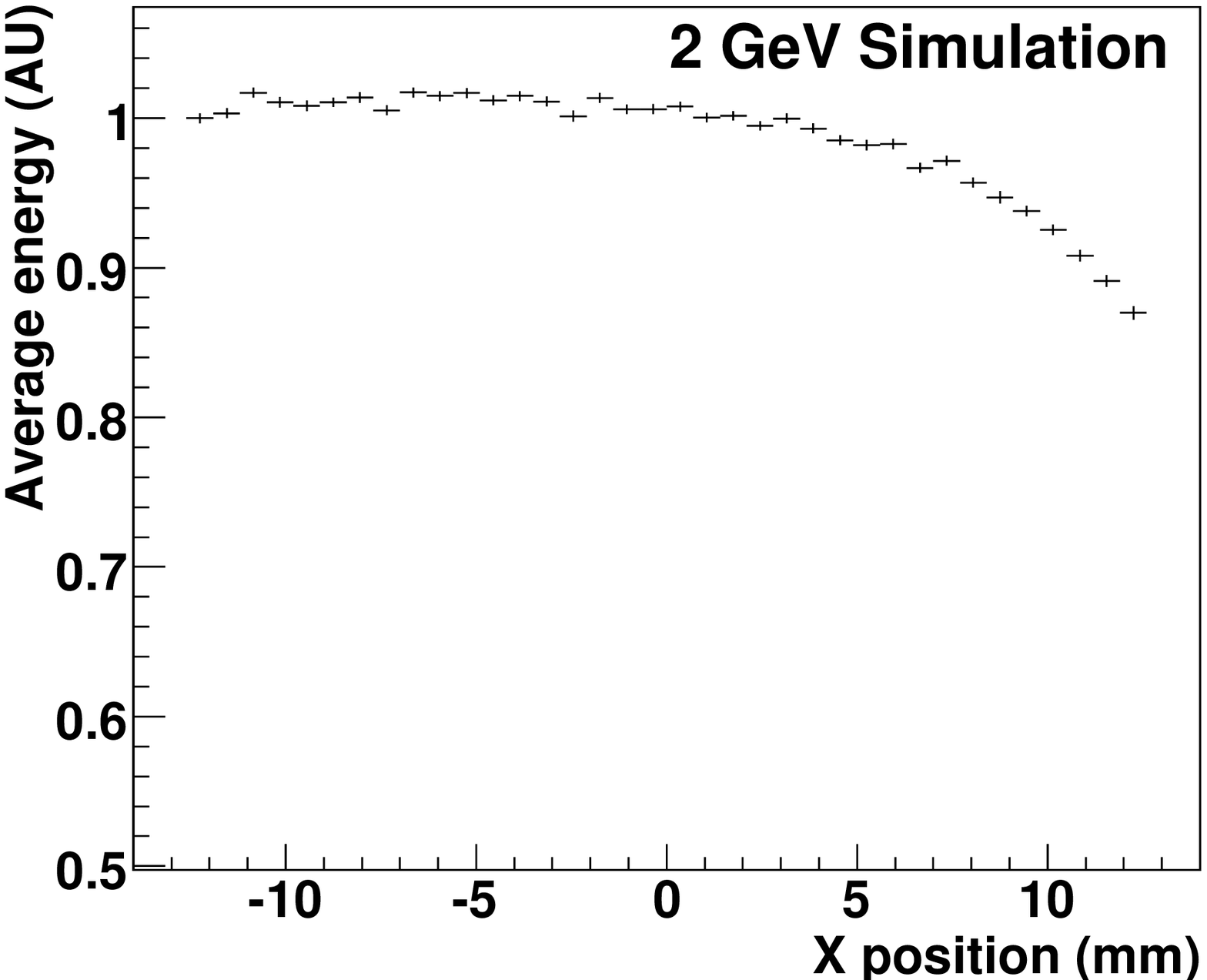}}

\end{centering}
\caption{a) Profiles of the average energy in a single module versus
inpact position of the beam for (a) data and (b) Monte Carlo.  The
sag toward the right is due to shower leakaage owing to the 5-degree
impact angle.} \label{fig:xprofiles}
\end{figure}

At FNAL, a broad negative beam was transported through a Cherenkov
counter (used to identify electrons), scintillator paddles, and
several multi-wire proportional chambers. Finally, the beam would
impact on the front faces of the W/SciFi prototype and two
$9\times9$~cm$^2$ Pb/SciFi ``reference'' detectors (each longer than
20~$X_0$ in depth).  Because the W/SciFi detector uses green fibers,
it was read out using a Hamamatsu R7899MOD (Extended Green
Photocathode, 1-inch diameter, Prismatic Window) photomultiplier
tube. Placed at each side of the tungsten detector, the Pb/SciFi
calorimeters were coupled to standard 2-inch bialkali PMTs.  These
detectors were used to provide shower containment for the tungsten
module, as well as to provide reference energy resolution
measurements under the same beam conditions.  Each of the three
calorimeters was wrapped in black electrical tape to keep them
individually lighttight.
The beamline wire chambers permitted horizontal and vertical impact
resolution with approximately 3-mm precision. The incident default
angle for both data taking and Monte Carlo was 5 degrees.
Figure~\ref{fig:xprofiles} shows the average energy in a module
versus beam impact position over a 25~mm centered impact range. The
dip in the average at larger horizontal impact values (for both data
and Monte Carlo) is due to leakage for incomplete shower containment
because the beam enters at a 5-degree impact angle.

\begin{figure}
\begin{centering}
\includegraphics[width=0.75\textwidth]{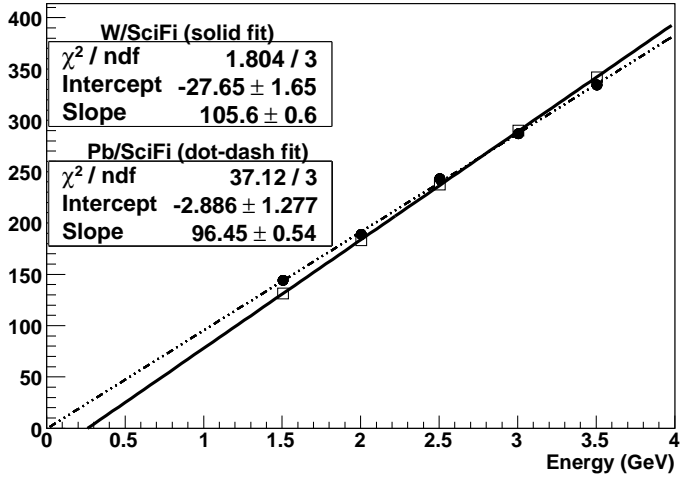}
\caption{Fitted mean versus beam energy for the W/SciFi (solid) and
one of the Pb/SciFi (dot-dashed) detectors with impact position cuts
to ensure contained showers. The linearity of the tungsten detector
is good; however, the evident negative intercept (-28 ADC counts)
was likely caused by a small dc offset that was integrated over the
gate width. One of the Pb/SciFi points pulls the chi-squared
goodness of fit.} \label{fig:linearity}
\end{centering}
\end{figure}

An energy sweep under the conditions that the beam impacted well within
the boundary of either the W/SciFi or one of the Pb/SciFi detectors leads
to the linearity plot shown in Fig.~\ref{fig:linearity}.  While the lead detector
has a vertical intercept consistent with zero, the tungsten detector has a
negative offset of 28 channels.  This offset was originally unnoticed, and is
presumed to be tied to a small positive dc bias integrated over a large
ADC gate.  The 28-channel offset is corrected for in the analysis that follows.

At PSI, a similar detector setup was used at the piM1 beamline.
Again, a broad negative beam was transported to the detectors. Small
scintillator paddles were used to trigger beam impact on the W/SciFi
within the edge boundaries.  The beam consisted of a small fraction
of electrons as compared to pions and muons.  Timing with respect to
the machine 50~MHz RF was used to select the electron signal and
out-of-time sideband subtraction was also used to remove unwanted
background.

\subsection{Energy Resolution} 
\begin{figure}
\begin{centering}
\includegraphics[width=0.75\textwidth]{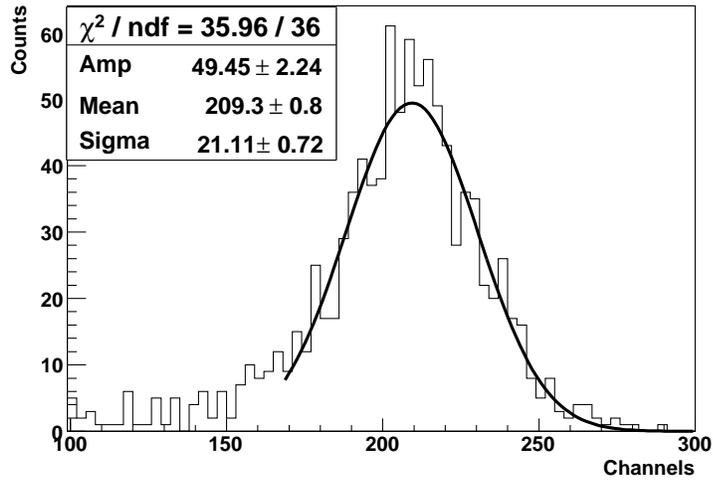}
\caption{Example raw W/SciFi detector ADC data for a 2~GeV electron
beam impacting at a 5 degree incidence. A modest containment cut of
15~mm width is made. } \label{fig:2gevres}
\end{centering}
\end{figure}

The typical \gm\ energy threshold for including events is 2~GeV;
positrons are only accepted above this threshold.  A resolution of
$\sim10\%$ (defined as $\sigma/E$ for a simple Gaussian fit) would
give an acceptable performance for this threshold in a future \gm\
experiment. The response of the prototype tungsten detector is shown
in Fig.~\ref{fig:2gevres} for 2~GeV electrons impacting at 5~degrees
with respect to the fiber axis.  An entrance cut is made using the
wire chamber information to select the central 15~mm by 30~mm (width
by height) of the 40- by 60-mm detector face. Even with no
corrections for leakage into side detectors, or adjustments for
sub-optimal light collection, or beam momentum uncertainty, (see
below) the resolution at $10.1\pm0.3\%$ meets the experimental goal.

\begin{figure}

\begin{centering}
\includegraphics[width=\textwidth]{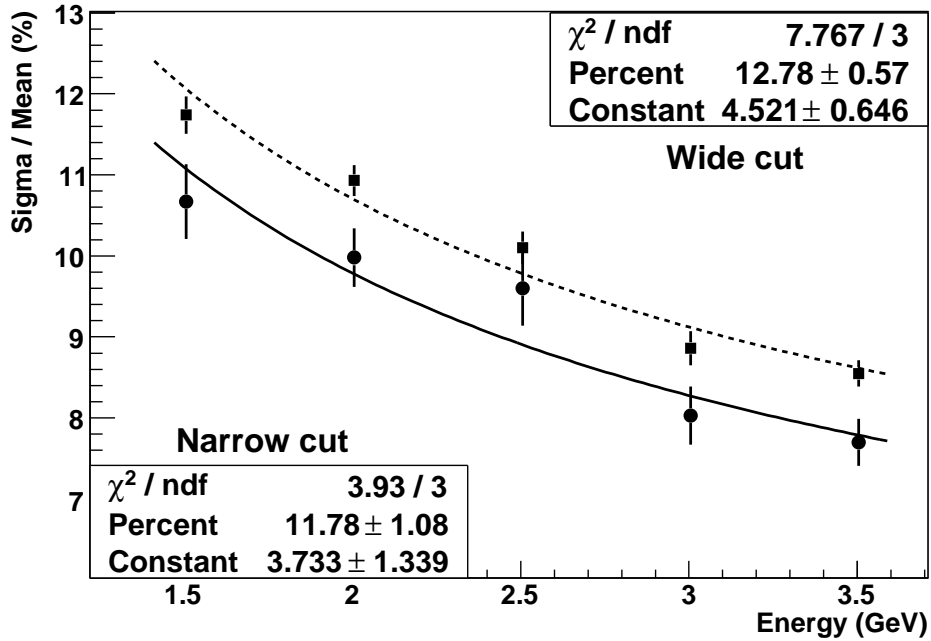}
\caption{Measured resolution at 5-degree impact angle versus energy.
The upper curve (dotted) is a fit to data obtained requiring a
25-mm-wide entrance cut.  The lower curve (solid) is a fit to data
obtained using a 5-mm-wide entrance cut. The``Percent'' term
represents the intrinsic sampling term ($A$ in the text); a $5\%$
photo-statistics stochastic term has been removed in the fit
function.\label{fig:resolution}}
\end{centering}
\end{figure}

Our goal is to understand the intrinsic sampling resolution of this
detector and compare it to simulation. While the stochastic term is
mainly determined by the sampling fluctuations intrinsic to the
active-to-absorber material ratio and the effective layer thickness,
additional contributions enter from photo-statistics.  A $5\%$
contribution exists from photo-statistics, because the measured
photoelectron (pe) yield is 400~pe/GeV.  This is a smaller light
yield than we would expect had the lightguide been better matched in
area to the photomultiplier tube and if blue fibers were used
instead of green (higher quantum efficiency). Two factors that scale
with energy contribute to energy degradation---the transverse
leakage, and the momentum uncertainty of the test beam. The leakage
can be explored with data cuts and simulation; the $\Delta P/P$ is
unknown but estimated to be a few percent. In
Fig.~\ref{fig:resolution}, we plot the FNAL data fit to
\begin{equation}
\frac{\sigma(E)}{E} = \sqrt{ \frac{A^2 + \Delta_{pe}^2}{E} + B^2 }.
\end{equation}
Here $A/\sqrt{E}$ represents the intrinsic sampling term,
$\Delta_{pe}/\sqrt {E}$ is the photo-statistics uncertainty, and $B$
is a linear term.  The term $\Delta_{pe}$ is fixed at $5\%$ and $E$
is given in GeV. The upper curve is a fit based on data where a
25~mm ``wide cut'' in the entrance width of the beam was used, while
the lower curve is a fit based on a 5~mm ``narrow cut.''  The change
affects both the sampling and the constant term as they are not
easily separable, given the statistics. The narrow cut result
minimizes, but does not eliminate, the leakage, resulting in
$A_{meas} = 11.8 \pm 1.1\%$ and $B_{meas} = 3.7 \pm 1.3\%$ for the
stochastic and constant terms, respectively. The same entrance cuts
could not be made with the three low-energy PSI data points owing to
the different setup, but the fit results did not change when those
points, which are dominated by the stochastic terms, were included.

\begin{figure}
\begin{centering}
\includegraphics[width=\textwidth]{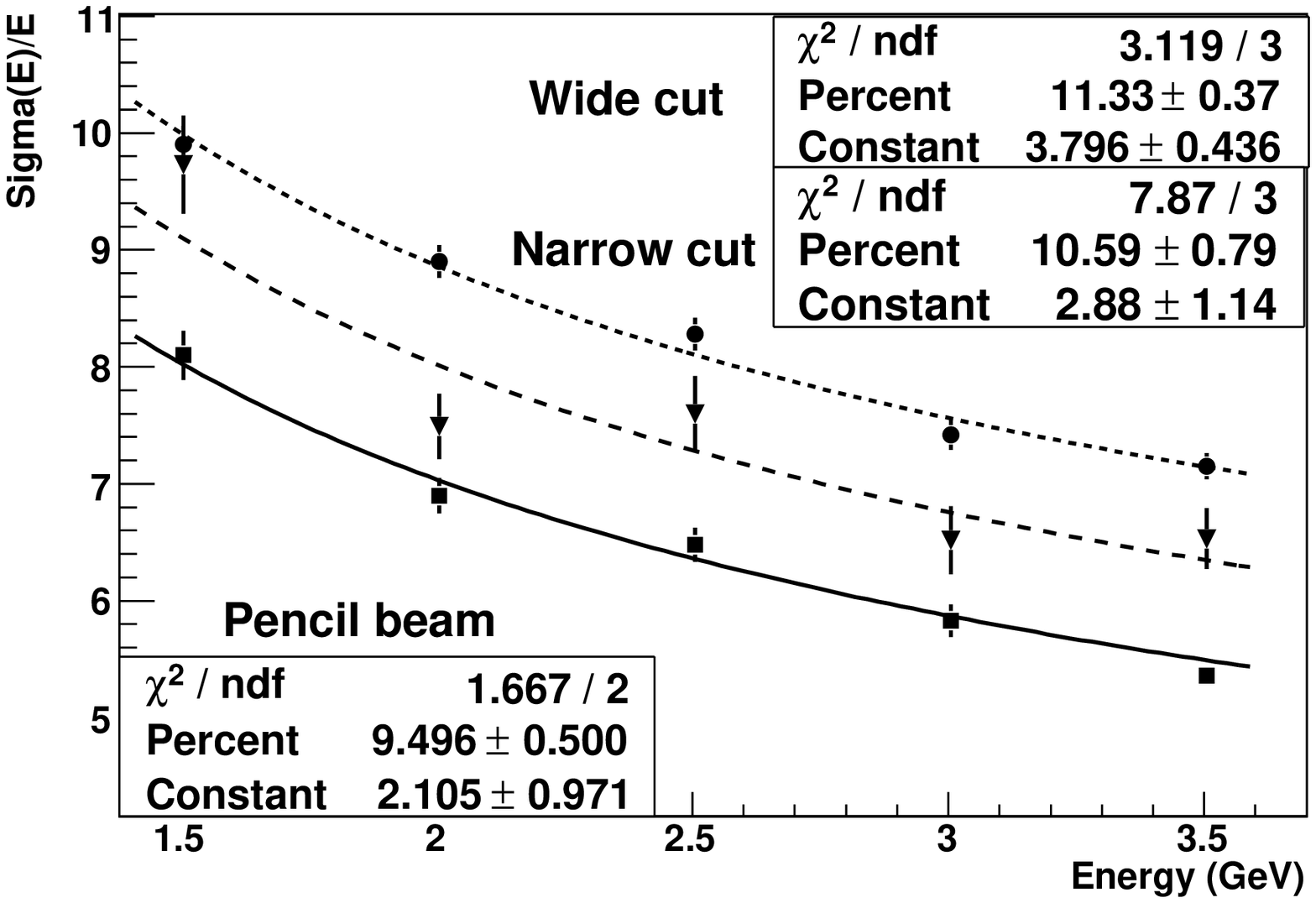}
\caption{Fits to resolution versus energy in the central module of
an array of W/SciFi modules. Three entrance width cuts are imposed:
25~mm (dotted), 5~mm (dashed), and 1~mm ``pencil'' (solid).
\label{fig:MCresolution}}
\end{centering}
\end{figure}

The sampling fluctuation component can be predicted using a complete
GEANT-4 model~\cite{geant}. A plot of this resolution versus energy
for simulated electrons impacting on the module center at a 5-degree
angle is shown in Fig.~\ref{fig:MCresolution}. Three curves are
presented representing a high-statistics ``pencil beam'' with a 1~mm
spot size in both dimensions, as well a separate simulation with
data-like cuts of 5- and 25-mm entrance widths, which match the
narrow and wide definitions for the data.  The most appropriate
comparison to data is the narrow cut, which yields $A_{sim} = 10.6
\pm 0.8\%$ and $B_{sim} = 2.9 \pm 1.1\%$ for the stochastic and
constant terms, respectively. The $B$ term is representative of the
leakage, since no $\Delta P/P$ uncertainty contributes for Monte
Carlo. If we deconvolute the leakage contribution from the $B$ term
in the data, a $\Delta P/P$ of $\approx 2.3\%$ is implied, which is
not unreasonable.

The simulation is, not surprisingly, somewhat better than the actual
prototype---$A_{sim} = 10.6 \pm 0.8\%$ vs. $A_{meas} = 11.8 \pm
1.1\%$. Detector construction imperfections can contribute, as would
non-uniform light collection in the guide.  However, to explore this
comparison more completely will require a larger test module with
improved readout and a better controlled test beam environment.
Note, that we carefully checked the GEANT-4 cut parameters, but
found no dependence on them that altered our results.

\subsection{Transverse and Longitudinal Profile}
\begin{figure}
\begin{centering}
\subfigure[Transverse Profile]
{\includegraphics[width=0.45\textwidth]{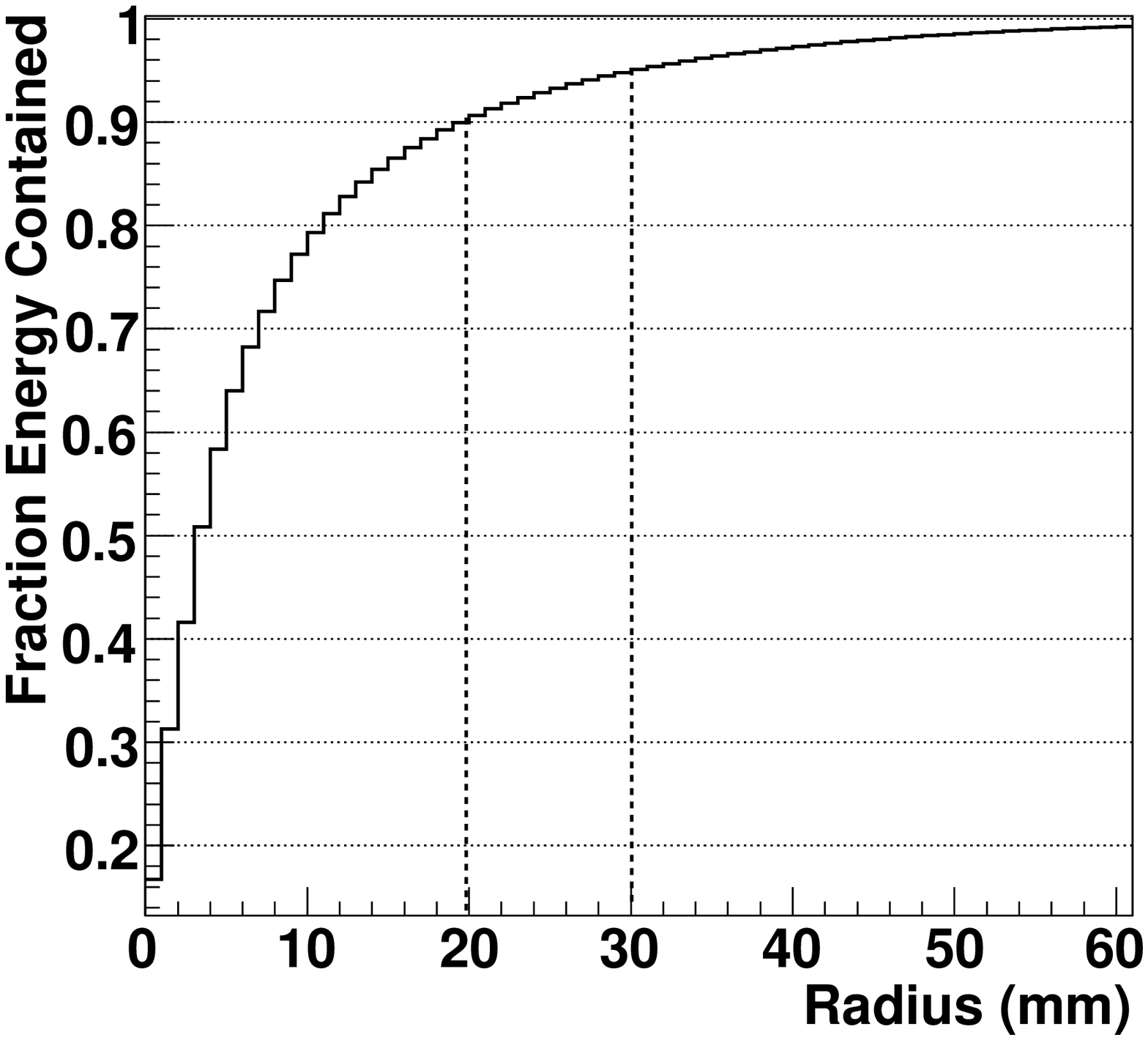}}
\subfigure[Shower Max vs Energy]
{\includegraphics[width=0.45\textwidth]{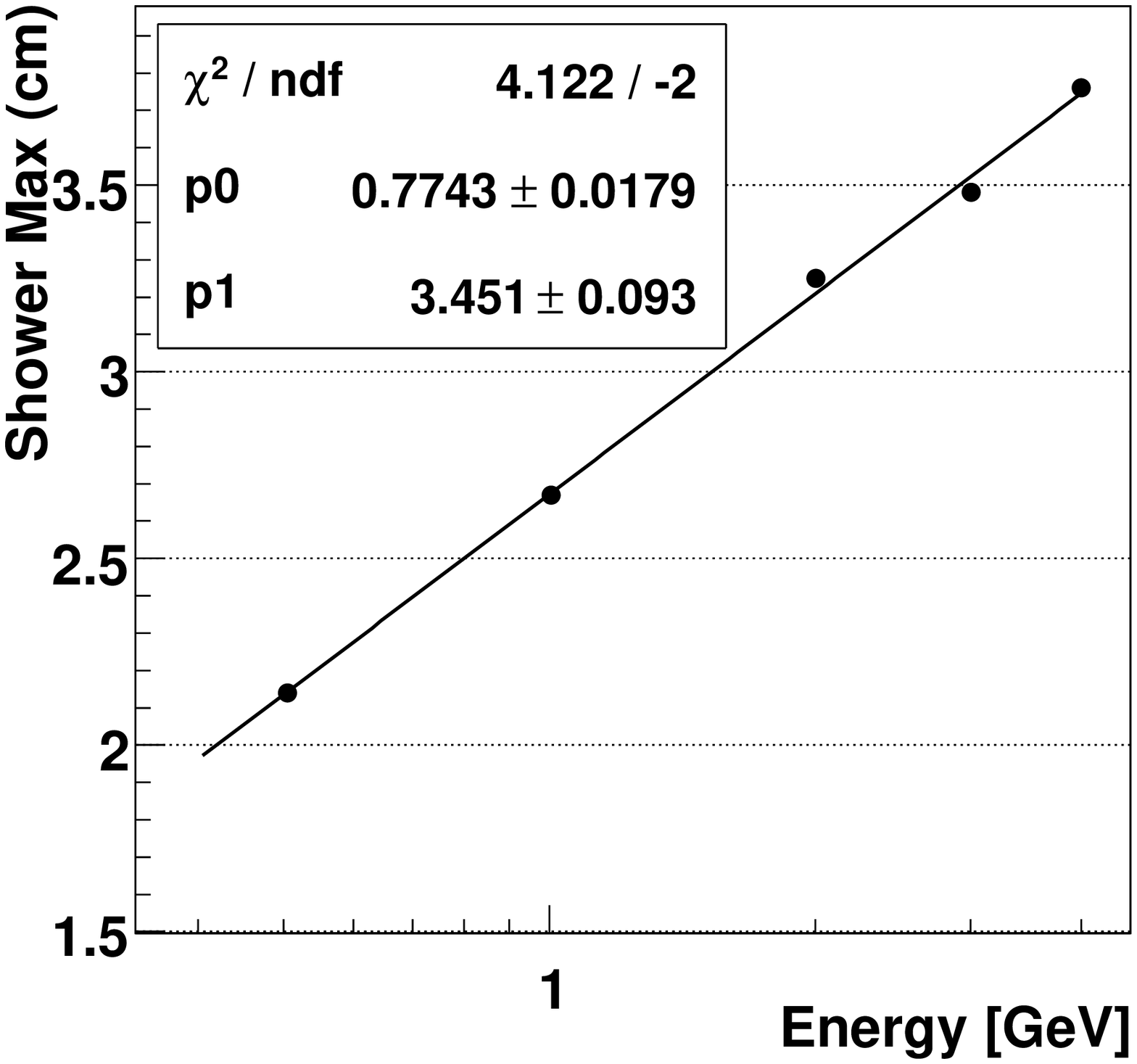}}

\end{centering}
\caption{Sample GEANT-4 simulations. (a) Integrated transverse
containment of 2~GeV electrons with $90\%$ and $95\%$ containment
radii of 20 and 30~mm indicated, respectively. (b) Linear fit of the
shower max from the longitudinal profile verses the log of energy.
The slope is the effective simulated radiation length, $X_0 =
0.77$~cm.  Both simulated figures of performance are larger than the
standard analytical expectations.} \label{fig:profiles}
\end{figure}

A GEANT-4 model was used to study the transverse and longitudinal
shower profiles and compare to the standard analytical
calculations~\cite{fabjan}. Energy containment as a function of
transverse distance for a pencil beam is shown in
Fig.~\ref{fig:profiles}a. The vertical dotted lines indicate the
radii of $90\%$ and $95\%$ containment, respectively.  The $90\%$
line corresponds to the simulated Moli\`{e}re radius of
approximately $2.0$~cm, which is greater than the 1.73~cm
expectation from the calculation. The depth of the longitudinal
profile maximum---shower max---increases proportional to $\ln(E)$;
typically $x = X_0\ln(E) + const$, where $x$ is the shower maximum.
The profiles of energy deposit versus depth for a range of incident
energies were fit to determine the shower maxima.
Figure~\ref{fig:profiles}b shows a plot of shower max versus energy.
A linear relation is evident with the energy axis plotted on a log
scale.  The slope of the fit corresponds to the simulated radiation
length, determined to be $X_0 = 0.77$~cm, which is also somewhat
larger than the analytical expectation of 0.69~cm.

\subsection{Angle of Impact on Layered Sampling Calorimeters}
\begin{figure}
\begin{centering}
\includegraphics[width=0.75\textwidth]{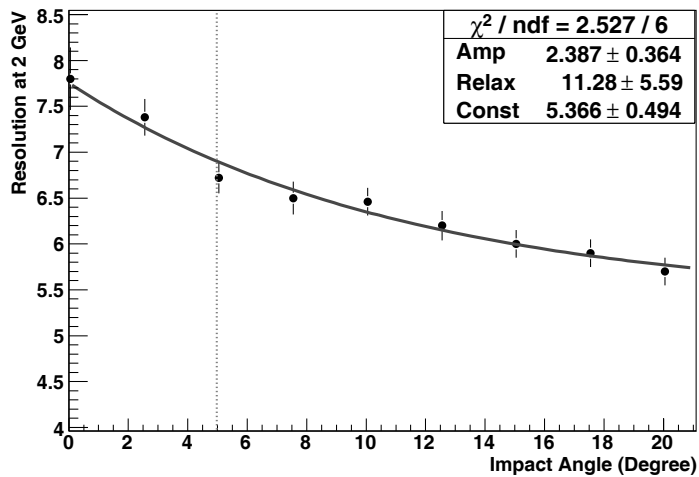}
\caption{GEANT-4 study of resolution for 2~GeV electrons versus
impact angle.  The points are fitted to $\sigma(\theta) = A
exp(-\theta/R)+B$, where $R$ is a relaxation term and $B$ is the
resolution at 90 degrees. The vertical dotted line represents the
minimal 5-degree impact proposed for a future \gm\ experiment.}
\label{fig:resvsangle}
\end{centering}
\end{figure}

It is possible to orient a lead- or tungsten-scintillating
calorimeter at near head-on impact with respect to the incoming
beam.  This permits readout at the rear, as well as near arbitrary
transverse segmentation.  As zero degree orientation is approached,
channeling---long particle paths through the low-$Z$ active
material---can occur without showering, which reduces the resolution
and the containment fraction.  The layers appear ``thicker'' at
shallow angles since the secondaries in the original shower
development largely retain the incoming electron direction (only at
the end are they more isotropic).  For \gm, the impact angle on a
calorimeter face varies from a few degrees to $\sim30$ degrees for
positrons of interest. It is possible to tilt the calorimeter
outward so that 5~degrees represents a minimum striking angle, which
is the motivation for this impact choice in many of the studies
reported here. The expected improvement in resolution at larger
angles was simulated using 2~GeV incoming electrons spread uniformly
over one tungsten plate and fiber layer.  The result is shown in
Fig.~\ref{fig:resvsangle}.  A parameterization is given by a
function of the form: $\sigma(\theta) = A exp(-\theta/R)+B$, where
$R$ represents a relaxation term and the constant term $B$ is the
resolution at 90-degrees.

While the narrow width of the module and limited statistics
prevented an exhaustive test, a comparison at 2~GeV was made using
beam data at 5 and 15-degree impact and with a 15-mm-wide by 20-mm
high entrance cut centered on the module front face.  The fractional
resolution improved in the data by $\sim 12\%$, which is similar to
the reduction found in the Monte Carlo for the same angular impact
change (see Fig.~\ref{fig:resvsangle}).

\subsection{Resolution versus layer thickness}
\begin{figure}
\begin{centering}
\includegraphics[width=0.75\textwidth]{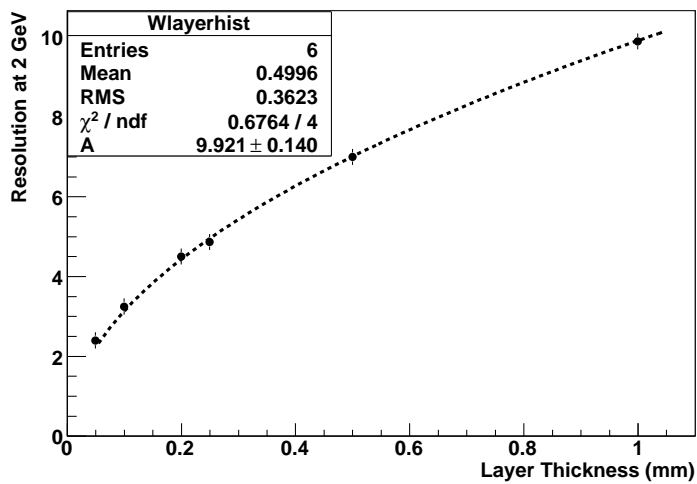}
\caption{GEANT-4 study of resolution for 2~GeV electrons versus
thickness of tungsten and fiber ribbon layers.  The fit function is
$\sigma(t) = A\sqrt{t}$, where $t$ is the thickness.}
\label{fig:layersweep}
\end{centering}
\end{figure}

Resolution improves for smaller effective layer thicknesses when
fixing the absorber-to-active material ratio.  This was one of the
driving motivations to develop scintillating-fiber-based
calorimeters~\cite{Hertzog:1990}. Commercial vendors have improved
fiber production quality and delivery options. Since there is a
larger range in sizes for calorimeter components, it is now worth
carefully evaluating the resolution-to-cost ratio for a given
application.  In Fig.~\ref{fig:layersweep}, we show a simulation for
5-degree impact 2-GeV electrons versus the thickness of the fiber
ribbons, while maintaining a constant 50:50 fiber/epoxy:tungsten
(volume) ratio. The overall module width is kept fixed at 40~mm and
the electron impact is centered across a horizontal width of twice
the layer thickness.  The figure includes a simple fit to $\sigma(t)
= A\sqrt{t}$, a good model that follows expectations for sampling
scaling with layer thickness $t$. Detector costs typically scale
with overall fiber length and plate number (linearly and not by
volume, for both product purchase and assembly considerations).
Resolution degradation terms (typically constant terms) increase
with detector imperfections, such as material tolerances and
construction imperfections. As is evident from
Fig.~\ref{fig:layersweep}, the intrinsic resolution only improves as
the square root of the thickness. Therefore, it is a matter of the
demands of the application and a delicate balance of constant terms
and photo-statistics limitations to determine where to set the layer
thicknesses. For our application, the 0.5-mm layer thickness is
close to ideal, but with improved light collection, perfected
assembly procedures, and strict mechanical tolerance specifications,
one could aim for a factor of 2 reduction in layer thickness.
However, fibers would have to be carefully tested for attenuation
losses at such a small diameter.

\section{Summary} 

A dense and fast electromagnetic calorimeter prototype has been
described.  With 0.5-mm layers of tungsten plates and scintillating
fiber ribbons, it promises acceptable resolution compared to the
Pb/SciFi detectors on which the concept is based, but at a
calculated radiation length of only 0.69~cm.  With fine-tuning of
the W/SciFi ratio or the layer thickness, resolution can be adjusted
to meet application demand. The straight-forward assembly procedure
can be scaled for modules of different widths and lengths.  The
small transverse shower dimensions ($R_M = 1.73$~cm) is appropriate
for use with small-area PMTs or with new SiPMs.

\ack{} 

Thanks to C. Polly for reading the manuscript, Y. Kuno for providing
the green-sensitive PMT, E. Ramberg for assistance with the FNAL
test beam, and P. Cooper for help with the FNAL wire chamber
tracking and positioning code. This work was supported by the United
States National Science Foundation.


\end{document}